\documentclass[journal]{IEEEtran}
\usepackage{amsmath}
\usepackage[center]{subfigure}
\usepackage{lineno,hyperref}
\usepackage{xcolor}
\usepackage{extarrows}
\usepackage{algorithm}
\usepackage{algorithmic}
\usepackage{setspace}
\usepackage{amssymb}
\usepackage{bm}
\usepackage{graphicx}
\usepackage[justification=centering]{caption}
\usepackage{cite}
\usepackage{subfigure}
\usepackage{url}
\usepackage{multirow}
\usepackage{tabularx}
\usepackage{booktabs}
\usepackage{tikz}
\usetikzlibrary{arrows,positioning,fit,shapes,calc,decorations.pathreplacing,matrix,patterns,quotes,angles}
\usepackage{tikz-3dplot}
\usepackage{cases}
\usepackage{arydshln}

\newcommand{\rev}{\textcolor{black}}

\newcommand{\larrow}{\overset{\scriptscriptstyle\leftarrow}}
\newcommand{\rarrow}{\overset{\scriptscriptstyle\rightarrow}}

\begin{document}

\title{Joint Near Field Uplink Communication and Localization Using Message Passing-Based Sparse Bayesian Learning} 

\author{ Fei Liu, Zhengdao Yuan, Qinghua Guo, \IEEEmembership{Senior Member, IEEE}, Yuanyuan Zhang,  Zhongyong Wang and J. Andrew Zhang, \IEEEmembership{Senior Member, IEEE}
	\thanks{ The work of F. Liu, Z. Yuan, Z. Wang and Y. Zhang was supported in part by National Natural Science Foundation of China (62101506) .} 
	\thanks{F. Liu and Z. Wang are with the School of Geoscience and Techonology and the School of  Electrical and Information Engineering, Zhengzhou University, Zhengzhou 450000, China, (e-mail: ieliufei@hotmail.com, zywangzzu@gmail.com)}
	\thanks{Z. Yuan is with the Artificial Intelligence Technology Engineering Research Center, Open University of Henan
		Zhengzhou 450000, China (e-mail: yuan\_zhengdao@163.com).}
	\thanks{Q. Guo is with the School of Electrical, Computer and Telecommunications Engineering, University of Wollongong, Wollongong, NSW 2522, Australia  (e-mail: qguo@uow.edu.au).}
	\thanks{Y. Zhang  is with the School of Electronics and Information, Zhengzhou University of Light Industry, Zhengzhou 450002, Henan, China. She was with the School of Electrical, Computer and Telecommunications Engineering, University of Wollongong, Wollongong, NSW 2522, Australia (e-mail: ieyyzhang@zzuli.edu.cn)}
	\thanks{J. A. Zhang is with the Global Big Data Technologies Centre, University of Technology Sydney, Sydney, NSW 2007, Australia (Email: Andrew.Zhang@uts.edu.au)}
}

\maketitle

\begin{abstract}
	This work deals with the problem of uplink communication and localization in an integrated sensing and communication system, where users are in the near field (NF) of antenna aperture due to the use of high carrier frequency and large antenna arrays at base stations.
	We formulate joint NF signal detection and localization as a problem of recovering signals with a sparse pattern.
	To solve the problem, we develop a message passing based sparse Bayesian learning (SBL) algorithm, where multiple unitary approximate message passing (UAMP)-based sparse signal estimators work jointly to recover the sparse signals with low complexity. Simulation results demonstrate the effectiveness of the proposed method.
	
\end{abstract}


\section{Introduction}
Large antenna arrays operating in high frequency bands such as millimeter and THz bands have attracted much attention for their potential applications in 6G \cite{cui2023, 9508850, liu2023}. Due to the use of large antenna aperture at base stations (BSs) and high frequency bands, mobile terminals often fall in the near field (NF) of array aperture as the Fraunhofer distance, which is the boundary between NF and far field (FF) regions, 
is usually large \cite{nftechrxiv, wei2022, wang2023}. For instance, we assume an {uniform linear antenna array with 64 elements (half wavelength spacing) and a carrier frequency of 6GHz, the Fraunhofer distance is about 99m.} Recently,  integrated sensing and communication (ISAC) in wireless networks has attracted much attention as sensing has been recognized as an integrated part of wireless networks, e.g., localization will be a service in wireless networks for many applications such as autonomous driving\cite{andrewZhangConcurrent}. In this work, we focus on joint NF communication and localization in  uplink transmission, where BSs perform the tasks of localization and signal detection.  

With the plane-wave assumption used for FF, a multitude of time delay and angle estimation algorithms such as multiple signal classification (MUSIC) based \cite{andrewZhangMultiple} and estimation of signal parameters via rotational invariance techniques (ESPRIT)\cite{32276} can be for user localization. However, in the NF region, the plane-wave assumption is no longer valid \cite{8736783, wei2022, cui2023}, and the methods developed with the plane-wave assumption suffer significant performance loss.  The works in \cite{primerNFBeforming, nfanalysis, liu2023, wei2022} consider NF channel modeling, estimation and system performance analysis, which, however, do not address the issue of NF localization. On the opposite, tracking with filter evaluations \cite{9508850}, and compressive sensing-based algorithm \cite{9709801} 
are proposed for NF localization. However, they only focus on NF localization and do not consider the issue of communications. A NF ISAC framework is proposed in \cite{wang2023}, where the ISAC waveform is optimized. 

In this work, we investigate  joint NF uplink communication and localization, where users transmit signals to BSs, and BSs locate the users and perform multi-user signal detection at the same time. To avoid the frequent use of pilot signals especially in the case of fast moving users, we propose to use differentiation modulation schemes so that pilot signals are potentially not required. To tackle this challenging problem of joint localization and communication without using pilots, we formulate it as a problem of recovering sparse signals with sparse pattern. To solve the problem with low complexity, we design a message passing based method, where multiple unitary message passing (UAMP) \cite{Guo2015UtAMP} based sparse signal estimators work jointly to recover the sparse signal. Simulation results demonstrate the effectiveness of the proposed method.

The remainder of the paper is organized as follows. In Section II, we introduce the system model and formulate the problem of joint NF uplink communication and localization. In Section III, the problem is reformulated as probabilistic form, and efficient message passing algorithms are developed leveraging a factor graph representation for the problem and UAMP. Simulation results are provided in Section IV to demonstrate the superior performance of the proposed algorithms, and conclusions are drawn in Section V. 

\textit{Notations}- Boldface lower-case and upper-case letters denote column vectors and matrices, respectively. 
The distribution of a complex Gaussian variable with mean {$\hat x$} and variance $\nu_x$ is represented by $\mathcal{CN}(x;{\hat x},\nu_x)$. We use $\text{diag}(\boldsymbol{z})$ to  represent a diagonal matrix with $\boldsymbol{z}$ as its diagonal. {The notations} $\boldsymbol{a}\cdot \boldsymbol{b}$ and $\boldsymbol{a} ./ \boldsymbol{b}$ represent the element-wise product and division between vector $\boldsymbol{a}$ and $\boldsymbol{b}$, respectively.
We use $\boldsymbol{1}$ and $\boldsymbol{0}$ to denote an all-one vector and an all-zero vector with a proper length, respectively. 
The notation $\text{vec}(\boldsymbol{X})$ represents the vectorization of matrix $\boldsymbol{X}$ column by column.
In addition, $q_j$ denotes the $j$-th entry of $\boldsymbol{q}$.

\begin{figure}[htbp]
	\centering
	\includegraphics[width=0.8\columnwidth]{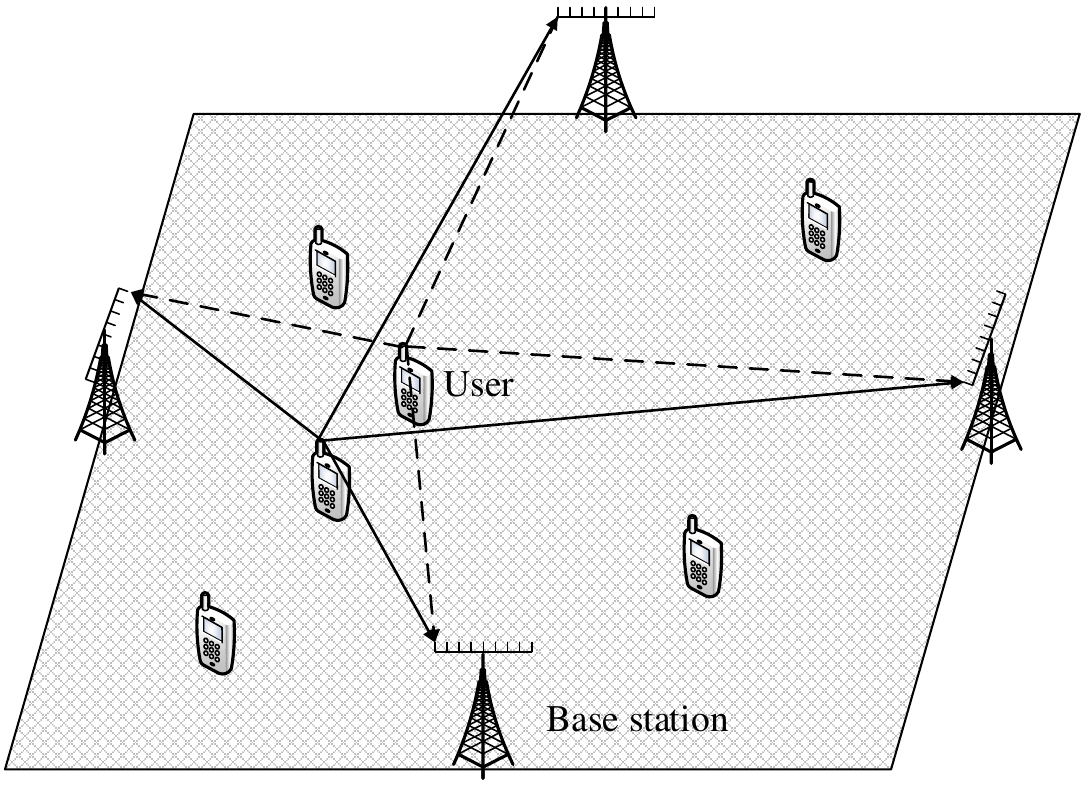} 
	\label{fig:sysmodel}
	\caption{{Illustration of the ISAC system. }} \label{fig:sysmodel}
\end{figure}

\section{{System Model and Problem Formulation}}  \label{sec:model}

As shown in Fig. 1, we consider an area covered by $L$ BSs, which serve $K$ users in the area. Assume that each BS is equipped with a large antenna array with $R$  elements, while each user is equipped with a single antenna. Although we do not have particular requirements on the array configuration, for simplicity, a uniform linear array (ULA) is considered in this work.
We consider uplink transmission, and the BSs collaboratively perform multi-user detection and user localization.

It is noted that the conventional plane-wave assumption is on longer applicable in NF \cite{NFSource}, thus the NF array response is introduced for modeling. Taking the $l$-th BS as example, where the first antenna is used as reference, the array response at the $r$-th antenna can be expressed as {\cite{NFChEstPosition} 
	\begin{align}
		a_{l,r}^k = \exp\left(-j\frac{2\pi}{\lambda}(d_{l,r}^k-d_{l,1}^k)\right),
	\end{align}
	where $\lambda$ denotes the signal wavelength, $d_{l,1}^k$ and $d_{l,r}^k$ denote the distance from user $k$ to the $1$-st and $r$-th antennas, respectively. The received signal from the $K$ users at the $l$-th base station can be expressed as 
	\begin{align}
		\boldsymbol{y}_l = \sum\nolimits_{k=1}^K g_l^k \boldsymbol{a}_l^k s_k + \boldsymbol{v}_l, \label{eq:Yl}
	\end{align}
	where $\boldsymbol{a}_l^k = [a_{l,1}^k, a_{l,2}^k, \cdots, a_{l,r}^k, \cdots, a_{l,R}^k]^T$, $\boldsymbol{y}_l \in \mathbb{C}^{R\times 1}$,  $g_l^k$ represents the channel coefficient \footnote {The unknown phase due to the imperfect synchronization between the user and BS can be absorbed into the channel coefficient.}, $s_k$ denotes the transmitted symbol of user $k$ and is drawn from the consellation set $\mathcal{A}$, $\boldsymbol{v}_l \in \mathbb{C}^{R\times 1}$ denotes the additive white Gaussian noise vector with mean zero and covariance matrix $\gamma^{-1}\boldsymbol{I}$. Collecting the received signals of all $L$ BSs and stacking them into one vector, we have
	\begin{align}
		\boldsymbol{y} = \sum\nolimits_{k=1}^K \boldsymbol{A}_k\boldsymbol{x}_k + \boldsymbol{v} \label{eq:ymodel}
	\end{align}
	where $\boldsymbol{y} = \left[\boldsymbol{y}_1^T, \boldsymbol{y}_2^T, \cdots, \boldsymbol{y}_L^T\right]^T \in \mathbb{C}^{RL\times 1}$, 
	$\boldsymbol{x}_k = s_k\left[g_1^k, g_2^k, \cdots, g_L^k\right]^T \in \mathbb{C}^{L\times 1}$, $\boldsymbol{v} = \left[\boldsymbol{v}_1^T, \boldsymbol{v}_2^T, \cdots, \boldsymbol{v}_L^T\right]^T \in \mathbb{C}^{RL\times 1}$ and 
	\begin{align}
		\boldsymbol{A}_k=\begin{bmatrix}
			\boldsymbol{a}_1^k & \boldsymbol{0} & \boldsymbol{\cdots} & \boldsymbol{0} \\
			\boldsymbol{0} & \boldsymbol{a}_2^k & \boldsymbol{\cdots} &  \boldsymbol{0}\\		
			\boldsymbol{\vdots} & \boldsymbol{\vdots} & \boldsymbol{\ddots} &  \boldsymbol{\vdots}\\	
			\boldsymbol{0} & \boldsymbol{0} & \boldsymbol{\cdots} &  \boldsymbol{a}_L^k\\
		\end{bmatrix}.\label{eq:Ak}
	\end{align}
	The above model is for a certain time instant, where each user transmits a single symbol. Considering a sequence of $N$ symbols, $N \in \mathbb{N}^{+} $, we have the following model
	\begin{align}
		[\boldsymbol{y}^1, \boldsymbol{y}^2, \cdots, \boldsymbol{y}^N] = \boldsymbol{A}[\boldsymbol{x}^1, \boldsymbol{x}^2, \cdots, \boldsymbol{x}^N] +  [\boldsymbol{v}^1, \boldsymbol{v}^2, \cdots, \boldsymbol{v}^N],  \label{eq:nymodel}
	\end{align}
	where { $\boldsymbol{A} =[\boldsymbol{A}_1, \boldsymbol{A}_2, \cdots, \boldsymbol{A}_K] \in \mathbb{C}^{RL \times KL}$, $\boldsymbol{x}^n = \left[ s_1^n\left[g_1^1, \cdots, g_L^1\right], \cdots, s_K^n\left[g_1^K, \cdots, g_L^K\right]\right]^T\in \mathbb{C}^{KL \times 1}$},   $\boldsymbol{y}^n$ and $\boldsymbol{v}^n$ denote the corresponding received signal vector and noise vector, respectively. 
	The aim is to locate all users and recover their transmitted symbols, i.e., both $\{\boldsymbol{A}_k\}$ and {$\{s_k^n\}$} need to be recovered. 
	
	A way that to circumvent the problem of unknown matrix $\boldsymbol{A}$ in model \eqref{eq:nymodel} is to transform it to a compressive sensing problem by dividing the covered area with a two dimensional uniform grid. 
	Suppose that the number of grid points is $M$, and a grid point with index $m$ corresponds to a matrix $\boldsymbol{A}_m$. Then we equivalently have 
	\begin{align}
		\boldsymbol{Y} =&\boldsymbol{\overline{A}}~\overline{\boldsymbol{X}} + \boldsymbol{V}, 
		\label{eq:sparseymodel}
	\end{align}
	where $\boldsymbol{Y} = [\boldsymbol{y}^1, \boldsymbol{y}^2, \cdots, \boldsymbol{y}^N] $, $\overline{\boldsymbol{A}} = [\boldsymbol{A}_1, \boldsymbol{A}_2, \cdots, \boldsymbol{A}_M] \in \mathbb{C}^{RL\times ML}$, $\overline{\boldsymbol{X}}= [\overline{\boldsymbol{x}}^1, \overline{\boldsymbol{x}}^2, \cdots, \overline{\boldsymbol{x}}^N] \in \mathbb{C}^{ML\times N}$ with  $\overline{\boldsymbol{x}}^n = \left[s_1^n\left[g_1^1, \cdots, g_L^1\right], \cdots, s_M^n\left[g_1^M, \cdots, g_L^M\right]\right]^T\in \mathbb{C}^{ML \times 1}$, and $\boldsymbol{V} = [\boldsymbol{v}^1, \boldsymbol{v}^2, \cdots, \boldsymbol{v}^N] \in \mathbb{C}^{RL\times N}$ denotes a Gaussian white noise matrix. As the user number $K$ is much smaller than $M$, the matrix $\overline{\boldsymbol{X}}$ is sparse . In particular, the column vectors in $\overline{\boldsymbol{X}}$ are block sparse (with block length $L$) and they share a common support. Based on the estimated locations of the nonzero blocks, we can determine the locations of the users. 
	
	Note that $\overline{\boldsymbol{X}}$ depends on {$\{s_m^n\}$ and $\{g_l^m\}$}, both of which are unknown, making the estimation of $\{s_m^n\}$ difficult. To avoid this issue, we propose to use differential modulations, so that once the estimate of $\overline{\boldsymbol{X}}$ is obtained, the transmitted symbols can be detected by performing differential demodulation without the knowledge of {$\{g_l^m\}$}. The advantage is that pilot symbols are no longer required, making the scheme more suitable for handling fast moving users. Regarding the formulated problem, we note that 
	\begin{itemize}
		\item The column vectors in matrix $\overline{\boldsymbol{X}}$ are block sparse, and they share a common support. The structure will be fully exploited in our proposed algorithm.
		\item Considering the size of matrix $\overline{\boldsymbol{A}}$, i.e., $RL\times ML$, low complexity sparse signal recovery algorithms are crucial.  
	\end{itemize}

	\section{Message Passing Algorithm Design for Joint Communication and Localization}
	
	A straightforward way to recover the matrix $\overline{\boldsymbol{X}}$ is to use the SBL algorithm \cite{Tipping2001Sparse}. However, it requires a complexity of $\mathcal{O}(M^3L^3 N)$ per iteration, which is a concern. In addition, the sparse pattern of $\overline{\boldsymbol{X}}$ cannot be exploited. To avoid these issues, we first develop a message passing-based Bayesian algorithm with much lower complexity to locate the users, and then symbol detection algorithm is designed with the knowledge of  users' locations.
	
	\begin{figure*}[htbp]
	\centering
	\includegraphics[width=1.9\columnwidth]{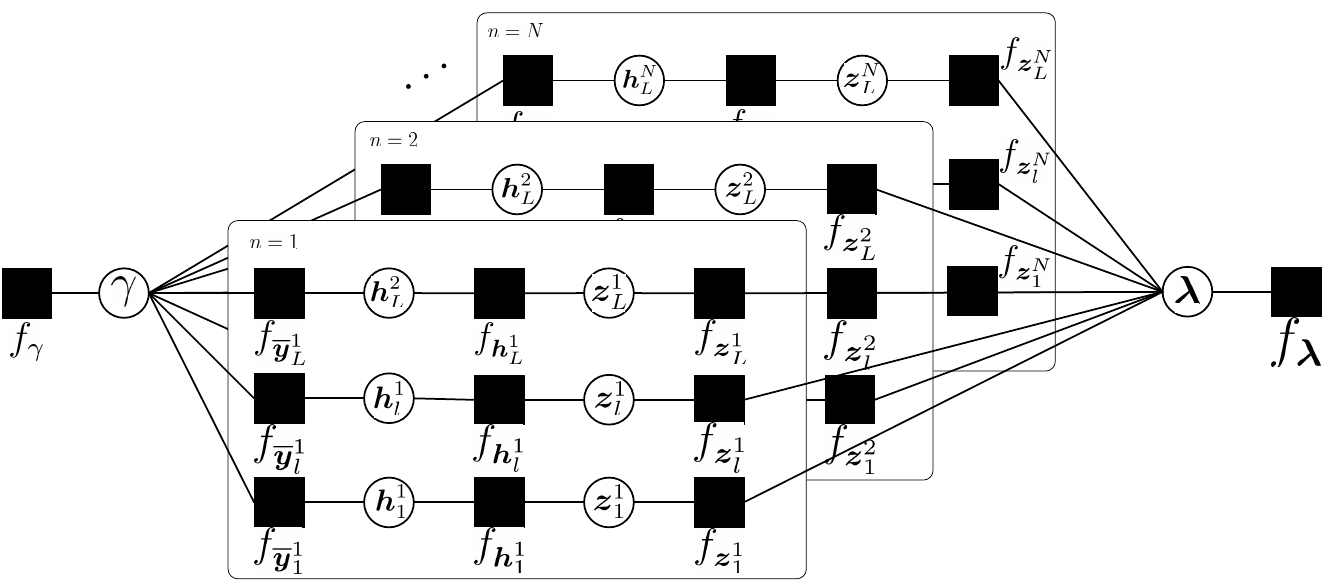} 
	\label{fig:nfcfg}
	\caption{{Factor graph representation of \eqref{eq:nfcfactor}}} 
\end{figure*}

	\subsection{Model Reformulation and Probabilistic Representation} 
	
	By exploiting the structure of matrix $\overline{\boldsymbol{A}}$, we can rearrange its columns properly to form a new block diagonal matrix 
	\begin{equation}
		\boldsymbol{H}=\text{diag}(\boldsymbol{H}_1, \boldsymbol{H}_2, \cdots,  \boldsymbol{H}_L ) \in \mathbb{C}^{RL\times ML}
	\end{equation}	
	where $\boldsymbol{H_l} = [\boldsymbol{a}_l^1, \boldsymbol{a}_l^2,\cdots , \boldsymbol{a}_l^M] \in \mathbb{C}^{R\times M}$. {Accordingly, we rearrange the entries of each column vector $\overline{\boldsymbol{x}}^n$ to form a new vector  
		\begin{align}
			\boldsymbol{z}^n = [(\boldsymbol{z}_1^n)^T, (\boldsymbol{z}_2^n)^T, \cdots, (\boldsymbol{z}_L^n)^T]^T, \label{eq:zn}
		\end{align}
	where $\boldsymbol{z}_l^n \in \mathbb{C}^{M\times 1}$ consists of the $\{l, l+L, \cdots, l+(M-1)L\}$-th entries of $\bar {\boldsymbol{x}}^n$, and then we obtain
		\begin{align}
			\boldsymbol{Y}= \boldsymbol{H}[\boldsymbol{z}^1, \boldsymbol{z}^2, \cdots, \boldsymbol{z}^N] + \boldsymbol{V}. \label{eq:Hmodel}
		\end{align}
		
		{Regarding the new model \eqref{eq:Hmodel}, we note that:
			\begin{itemize}
				\item Although \eqref{eq:Hmodel} is equivalent to the original model \eqref{eq:sparseymodel}, it allows us to design a sparse signal recovery algorithm by dealing with the much smaller matrices $\{\boldsymbol{H}_1, \boldsymbol{H}_2, \cdots, \boldsymbol{H}_L\}$, instead of the large matrix $\overline{\boldsymbol{A}}$. By leveraging UAMP, the computational complexity can be reduced drastically from $\mathcal{O}(M^3L^3 N)$ to $\mathcal{O}(RMLN)$. 
				\item Different from $\bar{\boldsymbol{x}}^n$, $\boldsymbol{z}^n$ is no longer block sparse. Instead, the sub-vectors $\{\boldsymbol{z}_1^n, \boldsymbol{z}_2^n, \cdots, \boldsymbol{z}_L^n\} $ are sparse and they share a common support. In addition, the vectors $\{\boldsymbol{z}^1, \boldsymbol{z}^2, \cdots, \boldsymbol{z}^N\}$ also share a common support.  
		\end{itemize}}
		
		To capture the sparse pattern mentioned above, we use the sparsity inducing hierarchical Gaussian-Gamma prior, i.e., 
		\begin{align}
			p(\boldsymbol{z}_l^n) = & \mathcal{CN}(\boldsymbol{z}_l^n; \boldsymbol{0}, \rev{\text{diag}(\{\lambda_1^{-1}, \lambda_2^{-1}, \cdots, \lambda_M^{-1}\}})), 	\\
			p(\lambda_m) = & Ga(\lambda_m; \epsilon, \eta).
		\end{align}
		As the noise variance is normally unknown, its estimation is also need to be considered. Thus, we assume that the precision of the noise {$\gamma$} has an improper prior $p(\gamma)=\gamma^{-1}$\cite{Tipping2001Sparse, ZHANG2017344}.

		To  facilitate the use of UAMP \cite{Guo2015UtAMP} on \eqref{eq:Hmodel} for joint communication and localizaitoin, we firstly perform unitary transformation to each block matrix i.e., $\boldsymbol{H}_l = \boldsymbol{U}_l\boldsymbol{\Lambda}_l\boldsymbol{V}_l, l=1,2,\cdots, L$, yielding
		\begin{equation} 
			\begin{aligned}
				\overline{\boldsymbol{y}}_l^n = \boldsymbol{\Phi}_l \boldsymbol{z}_l^n + \overline{\boldsymbol{v}}_l^n,~
				l=1,2,\cdots, L, 
				n=1, 2, \cdots, N,
			\end{aligned}
		\end{equation}
		where $\boldsymbol{\Phi}_l = \boldsymbol{\Lambda}_l\boldsymbol{V}_l$,  $\overline{\boldsymbol{y}}_l^n = \boldsymbol{U}_l^H\boldsymbol{y}_l^n$ and $\overline{\boldsymbol{v}}^n = \boldsymbol{U}_l^H\boldsymbol{v}_l^n$ with $\boldsymbol{y}_l^n$ and $\boldsymbol{v}_l^n$ denoting the $l$-th $R$-length sub-vector of $\boldsymbol{y}^n$ and $\boldsymbol{v}^n$, respectively. Define an auxiliary variable $\boldsymbol{h}_l^n \triangleq \boldsymbol{\Phi}_l\boldsymbol{z}_l^n$, then we have the following joint conditional distribution of variables $\{\boldsymbol{h}_l^n\}$, $\{\boldsymbol{z}_l^n\}$, $\{\lambda_m\}$ and $\gamma$,
		\begin{align}
			&p(\{\boldsymbol{h}_l^n\}, \{\boldsymbol{z}_l^n\}, \{\lambda_m\},\gamma | \{\overline{\boldsymbol{y}}_l^n\}) \nonumber \\
			=& \prod_{n,l}p(\overline{\boldsymbol{y}}_l^n|\boldsymbol{h}_l^n,\gamma)p(\boldsymbol{h}_l^n|\boldsymbol{z}_l^n)p(\boldsymbol{z}_l^n|\{\lambda_m\})p(\gamma)\prod_mp(\lambda_m)  \nonumber\\
			\triangleq & \prod_{n,l}f_{\overline{\boldsymbol{y}}_l^n}(\boldsymbol{h}_l^n,\gamma)f_{\boldsymbol{h}_l^n}(\boldsymbol{z}_l^n)f_{\boldsymbol{z}_l^n}(\{\lambda_m\})f_{\gamma}(\gamma)\prod_mf_{\lambda_m}(\lambda_m)		\label{eq:nfcfactor}
		\end{align}
		
		A factor graph representation of \eqref{eq:nfcfactor} is depicted in Fig. 2, in which the factors and the corresponding  distribution functions are listed in Table \uppercase\expandafter{\romannumeral1}. Since the message passing based alogrithm is usually iterative, where each iteration involves a forward (from left to right) and a backward (from right to left) message computation, we use $m_{a\rightarrow b}(x)$ to denote a message passed from a function node $a$ to a variable node $b$, which is a function of $x$, and use $n_{b\rightarrow a}(x)$ to denote a message passed from  variable node $b$ to a function node $a$, which is also a function of $x$. In addition, we use $b(x)$ to denote the belief of a varaible $x$. Note that, if a forward computation requires backword messages, the relevant messages in the previous iteration is used by default.
		
		\begin{table}[htb]
			\centering
			\renewcommand\arraystretch{1.2}
			\caption{Factors and distributions in (\ref{eq:nfcfactor}).}\label{tab:factor}
			\begin{tabular}{ccc}
				\hline
				Factor & Distribution & Function  \\
				\hline
				$f_{\overline{\boldsymbol{y}}_l^n}$&     $p(\overline{\boldsymbol{y}}_l^n|\boldsymbol{h}_l^n,\gamma)$ & $\mathcal{CN}(\overline{y}_l^n; \boldsymbol{h}_l^n, \gamma^{-1}\boldsymbol{I})$ \\	
				$f_{\boldsymbol{h}_l^n}$ & $p(\boldsymbol{h}_l^n|\boldsymbol{z}_l^n)$ & $\delta(\boldsymbol{h}_l^n - \boldsymbol{\Phi}_l\boldsymbol{z}_l^n)$\\
				$f_{\boldsymbol{z}_l^n}$ & $p(\boldsymbol{z}_l^n|\{\lambda_m\})$ & $\mathcal{CN}(\boldsymbol{z}_l^n; \boldsymbol{0}, \rev{\text{diag}(\{\lambda_m^{-1}\}}))$\\
				$f_{\gamma}$ & $p(\gamma)$ & $\gamma^{-1}$ \\		
				$f_{\lambda_m}$ & $p(\lambda_m)$ & $Ga(\lambda_m; \epsilon, \eta)$	\\	
				\hline
			\end{tabular}
		\end{table}						
		\vspace{-0.5cm}
		\subsection {Message Passing for Localization} \label{subsec:mplocalization} 
		In this subsection, we develop a message passing algorithm to obtain the approximate belief of the precisions  $\{\lambda_1, \lambda_2, \cdots, \lambda_M\}$, so that their estimates can be obtained, by which the locations of the users can be determined. With the aid of factor graph Fig.2, backward message computations and forward message computations are detailed in the following. 			
		
		\subsubsection{Backward Message Passing}
		Assume that the belief of $\lambda_m, 1\leq m \leq M$ is available and is denoted by $b(\lambda_m)$, as given later in \eqref{eq:belieflambdam}. Variational message passing (VMP) is used at the function node $f_{\boldsymbol{z}_l^n}(\boldsymbol{z}_l^n, \{\lambda_m\})$, and the message from $f_{\boldsymbol{z}_l^n}(\boldsymbol{z}_l^n, \{\lambda_m\})$ to the variable node $\boldsymbol{z}_l^n$ can be computed as
		\begin{align}
			m_{f_{\boldsymbol{z}_l^n}\rightarrow \boldsymbol{z}_l^n}(\boldsymbol{z}_l^n) &= \exp\left\{\int b(\{\lambda_m\})\ln f_{\boldsymbol{z}_l^n}(\boldsymbol{z}_l^n,\{\lambda_m\})d\{\lambda_m\} \right\}\nonumber \\ &\propto \mathcal{CN}(\boldsymbol{z}_l^n; \boldsymbol{0}, \text{diag}\{\rev{\hat{\lambda}_m^{-1}}\}), \label{eq:fznl2znl}
		\end{align}
		where $b(\{\lambda\}) \triangleq \prod_{m=1}^Mb(\lambda_m)$, $d\{\lambda_m\} \triangleq \prod_{m=1}^Md\lambda_m$, and $\hat{\lambda}_m$ denotes the estimate of $\lambda_m$, which is given later in \eqref{eq:lambdam}.
		
		The message $m_{f_{\boldsymbol{h}_l^n}\rightarrow \boldsymbol{z}_l^n}(\boldsymbol{z}_l^n)$ from function node $f_{\boldsymbol{h}_l^n}(\boldsymbol{h}_l^n, \boldsymbol{z}_l^n)$ to variable node $\boldsymbol{z}_l^n$ turns out to be Gaussian in \eqref{eq:fh2z} and $m_{f_{\boldsymbol{h}_l^n}\rightarrow \boldsymbol{z}_l^n}(\boldsymbol{z}_l^n)=\mathcal{CN}(\boldsymbol{z}_l^n; \boldsymbol{q}_l^n,\boldsymbol{\nu}_{\boldsymbol{q}_l^n})$. We can then compute the belief of $\boldsymbol{z}_l^n$ as
		\begin{align}
			b(\boldsymbol{z}_l^n)& = m_{f_{\boldsymbol{z}_l^n}\rightarrow \boldsymbol{z}_l^n}(\boldsymbol{z}_l^n)m_{f_{\boldsymbol{h}_l^n}\rightarrow \boldsymbol{z}_l^n}(\boldsymbol{z}_l^n) \nonumber \\
			& \propto \mathcal{CN}(\boldsymbol{z}_l^n;\hat{\boldsymbol{z}}_l^n,diag(\boldsymbol{\nu}_{\boldsymbol{z}_l^n})), \label{eq:beliefz}
		\end{align}
		where 
		\begin{align}
			\boldsymbol{\nu}_{\boldsymbol{z}_l^n} =& 1./(1./\boldsymbol{\nu}_{\boldsymbol{q}_l^n} + \rev{\hat{\boldsymbol{\lambda}}}), \label{eq:vznl}\\
			\hat{\boldsymbol{z}}_l^n =& \boldsymbol{\nu}_{\boldsymbol{z}_l^n} \cdot (\boldsymbol{q}_l^n./\boldsymbol{\nu}_{\boldsymbol{q}_l^n}), \label{eq:zhatnl}
		\end{align}
		and \rev{$\hat{\boldsymbol{\lambda}}=\text{diag} \{\hat{\lambda}_1, \hat{\lambda}_2, \cdots, \hat{\lambda}_M\}$}.
		
		Following UAMP, vectors $\boldsymbol{\nu}_{\boldsymbol{p}_l^n}$ and  $\boldsymbol{p}_l^n$ can be computed as
		\begin{align}
			\boldsymbol{\nu}_{\boldsymbol{p}_l^n} =& |\boldsymbol{\Phi}_l^n|^2\boldsymbol{\nu}_{\boldsymbol{z}_l^n}, \label{eq:vpnl} \\
			\boldsymbol{p}_l^n =& \boldsymbol{\Phi}_l^n\hat{\boldsymbol{z}}_l^n-\boldsymbol{\nu}_{\boldsymbol{p}_l^n}\cdot \boldsymbol{e}_l^n, \label{eq:phatnl}
		\end{align}
		where $\boldsymbol{e}_l^n$ is computed via \eqref{eq:eln} in the last iteration. According to the derivation of UAMP, we then have
		\begin{align}
			m_{f_{\boldsymbol{h}_l^n\rightarrow \boldsymbol{h}_l^n}}(\boldsymbol{h}_l^n) \propto \mathcal{CN}(\boldsymbol{h}_l^n;\boldsymbol{p}_l^n, \text{diag}(\boldsymbol{\nu}_{\boldsymbol{p}_l^n})).
		\end{align}
		With the message $m_{f_{\tilde{\boldsymbol{y}}_n^l\rightarrow \boldsymbol{h}_n^l}}(\boldsymbol{h}_n^l)$, whose computation is delayed to \eqref{eq:y2h}, the belief $b(\boldsymbol{h}_l^n)$ of $\boldsymbol{h}_l^n $ can be expressed as
		\begin{align}
			b(\boldsymbol{h}_l^n)& = m_{f_{\boldsymbol{h}_l^n\rightarrow \boldsymbol{h}_l^n}}(\boldsymbol{h}_l^n) m_{f_{\overline{\boldsymbol{y}}_l^n\rightarrow \boldsymbol{h}_l^n}}(\boldsymbol{h}_l^n) \nonumber \\
			& \propto \mathcal{CN}(\boldsymbol{h}_l^n; \hat{\boldsymbol{h}}_l^n, \text{diag}(\boldsymbol{\nu}_{\boldsymbol{h}_l^n})),
		\end{align}
		where
		\begin{align}
			\boldsymbol{\nu}_{\boldsymbol{h}_l^n} =& \boldsymbol{\nu}_{\boldsymbol{p}_l^n} ./(1+\hat{\gamma}\boldsymbol{\nu}_{\boldsymbol{p}_l^n}), \label{eq:vhnl} \\
			\hat{\boldsymbol{h}}_l^n =& \boldsymbol{\nu}_{\boldsymbol{h}_l^n}\cdot(\boldsymbol{p}_l^n./\boldsymbol{\nu}_{\boldsymbol{p}_l^n} + \overline{\boldsymbol{y}}_l^n\hat{\gamma}), \label{eq:hhatnl}
		\end{align}
		and $\hat{\gamma}$ is obtained in the last iteration with \eqref{eq:gammaest}
		
		Next, the message $m_{f_{\overline{\boldsymbol{y}}_l^n\rightarrow \gamma}}(\gamma)$ 
		is calculated with VMP, i.e.,
		\begin{align}
			m_{f_{\overline{\boldsymbol{y}}_l^n\rightarrow \gamma}}(\gamma) =& \exp \left\{\int b(\boldsymbol{h}_l^n)\ln f_{\widetilde{\boldsymbol{y}}_l^n}({\boldsymbol{h}}_l^n, \gamma) d\boldsymbol{h}_l^n\right\} \nonumber \\
			\propto &\gamma^{R}\exp \{-\gamma (||\overline{\boldsymbol{y}}_l^n - \hat{\boldsymbol{h}}_l^n||^2 + \boldsymbol{1}^T\boldsymbol{\nu}_{\boldsymbol{h}_l^n})\}.
			\label{eq:y2gamma}
		\end{align}
		This is the end of the backward message passing.
		
		\subsubsection{Forward Message Passing}
		According to VMP, we have
		\begin{align}
			m_{f_{\tilde{\boldsymbol{y}}_l^n\rightarrow \boldsymbol{h}_l^n}}(\boldsymbol{h}_l^n) &= \exp \left\{\int b(\gamma)\ln f_{\overline{\boldsymbol{y}}_l^n}({\boldsymbol{h}}_l^n, \gamma) d\gamma\right\} \nonumber\\ &\propto \mathcal{CN}(\boldsymbol{h}_l^n; \overline{\boldsymbol{y}}_l^n, \hat{\gamma}^{-1}\boldsymbol{I}_R), \label{eq:y2h}
		\end{align}
		where
		\begin{align}
			&b(\gamma) = f_{\gamma}(\gamma)\prod_{n=1}^N\prod_{l=1}^L m_{f_{\overline{\boldsymbol{y}}_l^n\rightarrow \gamma}}(\gamma) \nonumber \\ &\propto \gamma^{NLR-1}\exp\left\{-\gamma\sum_{n=1}^N\sum_{l=1}^L(||\overline{\boldsymbol{y}}_l^n - \hat{\boldsymbol{h}}_l^n||^2 + \boldsymbol{1}^T\boldsymbol{\nu}_{\boldsymbol{h}_l^n})\right\}
		\end{align}
		and the estimate of variable $\gamma$ can be expressed as
		\begin{align}
			\hat{\gamma} = \int \gamma b(\gamma) d\gamma  = \frac{NLR}{\sum_{N=1}^N\sum_{l=1}^L(||\overline{\boldsymbol{y}}_l^n - \hat{\boldsymbol{h}}_l^n||^2 + \boldsymbol{1}^T\boldsymbol{\nu}_{\boldsymbol{h}_n^l})}. \label{eq:gammaest}
		\end{align}
	
		Next, we update the intermediate vectors $\boldsymbol{\nu}_{\boldsymbol{e}_l^n}$  and $\boldsymbol{e}_l^n$ by
		\begin{align}
			\boldsymbol{\nu}_{\boldsymbol{e}_l^n} =& \boldsymbol{1}./(\boldsymbol{\nu}_{\boldsymbol{p}_l^n}+\hat{\gamma}^{-1}\boldsymbol{1}), \label{eq:veln}\\
			\boldsymbol{e}_l^n =& \boldsymbol{\nu}_{\boldsymbol{e}_l^n} \cdot (\overline{\boldsymbol{y}}_l^n - \boldsymbol{p}_l^n). \label{eq:eln}
		\end{align} 
		Then the message  $m_{f_{\boldsymbol{h}_l^n}\rightarrow \boldsymbol{z}_l^n}(\boldsymbol{z}_l^n)$ is computed as
		\begin{align}
			m_{f_{\boldsymbol{h}_l^n}\rightarrow \boldsymbol{z}_l^n}(\boldsymbol{z}_l^n) = \mathcal{CN}(\boldsymbol{z}_l^n; \boldsymbol{q}_l^n, \text{diag}(\boldsymbol{\nu}_{\boldsymbol{q}_l^n})), \label{eq:fh2z}
		\end{align}
		where
		\begin{align}
			\boldsymbol{\nu}_{\boldsymbol{q}_l^n} =& \boldsymbol{1}./\left(||\boldsymbol{\Phi}_l^n||^2\boldsymbol{\nu}_{\boldsymbol{e}_l^n}\right),\label{eq:vqnl} \\
			\boldsymbol{q}_l^n=& \hat{\boldsymbol{z}}_l^n + \rev{\boldsymbol{\nu}_{\boldsymbol{q}_l^n}} \cdot (\boldsymbol{\Phi}_l^n \boldsymbol{e}_l^n). \label{eq:qhatnl}
		\end{align}
		
		With the belief $b(\boldsymbol{z}_l^n)$ derived in \eqref{eq:beliefz}, the message $m_{f_{\boldsymbol{z}_l^n}\rightarrow \lambda_m}(\lambda_m)$ is calculated with VMP, i.e.,
		\begin{align}
			m_{f_{\boldsymbol{z}_l^n}\rightarrow \lambda_m}(\lambda_m) &= \exp\left\{\int b(z_{l,m}^n)\ln f_{z_{l.m}^n}(z_{l,m}^n,\lambda_m)dz_{l,m}^n\right\} \nonumber \\ &\propto \lambda_m \exp \left\{-\lambda_m\left(|\hat{z}_{l,m}|^2 + \nu_{z_{l,m}^n}\right)\right\}, 
			\label{eq:fz2lambda}
		\end{align}
		where $z_{l,m}^n$ denotes the $m$-th element of the vector $\boldsymbol{z}_l^n$. Then the belief of $\lambda_m$ can be computed as
		\begin{align}
			&b(\lambda_m) = f_{\lambda_m}(\lambda_m)\prod\nolimits_{n=1}^N\prod\nolimits_{l=1}^Lm_{f_{\boldsymbol{z}_l^n}\rightarrow \lambda_m}(\lambda_m) \nonumber\\ &\propto \lambda_m^{\epsilon-1+NL}\exp\left\{-\lambda_m \left(\eta + \sum_{n=1}^N \sum_{l=1}^L (|\hat{z}_{l,m}^n|^2 + \nu_{z_{l,m}^n})\right)\right\} \label{eq:belieflambdam}.
		\end{align}
		Then, the expectation of $\lambda_m$ can be updated as \cite{luomansbl}
		\begin{align}
			\hat{\lambda}_m &= \int \lambda_m b(\lambda_m)d\lambda_m \nonumber \\
			&= \frac{(\epsilon' + 1) NL}{\eta + \sum_{n=1}^N \sum_{l=1}^L \left( |\hat{z}_{l,m}^n|^2 + \nu_{z_{l,m}^n}\right)}, \label{eq:lambdam}
		\end{align}
		where $\epsilon'=\epsilon/NL$ and  $\epsilon'$ is updated with the following empirical rule \cite{luomansbl}
		\begin{align}
			\epsilon'= \frac{1}{2}\sqrt{\log \left(\frac{1}{M}\sum\nolimits_{m=1}^M \hat{\lambda}_m\right) - \frac{1}{M}\sum\nolimits_{m=1}^M\log\hat{\lambda}_m}. \label{eq:epsilon}
		\end{align} 
	}

	This is the end of the forward message passing. By performing the backward and forward message passing alternately until the normalized squared difference of the {estimated vectors $\{\boldsymbol{z}_l^n\}$} between two consecutive iterations is less than a threshold $\xi$, i.e., the algorithm converges, the estimated parameters $\{\lambda_m\}$ is used to determine the  locations of users by comparing it with a threshold $\beta$, i.e., if $\lambda_m < \beta$, we determine that a user is located at the $m$-th grid point.
	
%

	\subsection {Message Passing for Differential Demodulation} 	
		
	\begin{figure*}[htbp]
			\centering
			\includegraphics[width=1.4\columnwidth]{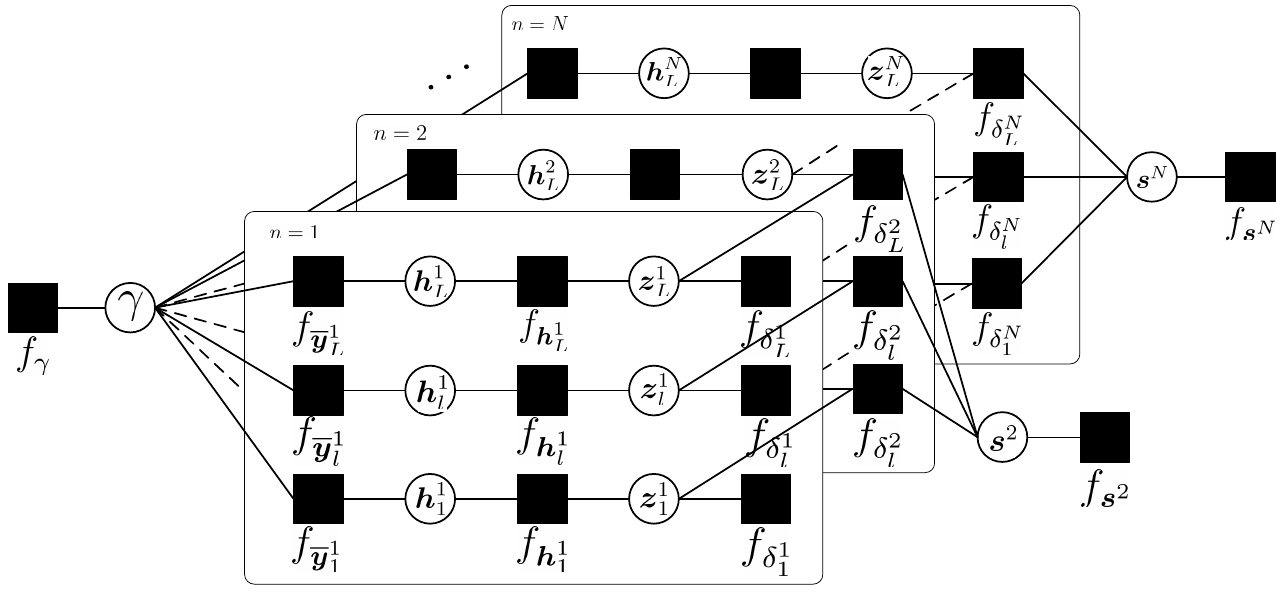} 			
			\caption{{Factor graph representation of \eqref{eq:comfactor} }} 
			\label{fig:stage2}
	\end{figure*}	
				
	With the result of users' localization, 
	the columns of $\boldsymbol{H}$ that correspond to zero entries in $\{\boldsymbol{z}_l^n\}$ and the zero entries of  $\{\boldsymbol{z}_l^n\}$ in \eqref{eq:Hmodel} and \eqref{eq:zn} can be removed, leading to a new matrix $\boldsymbol{H}' \in \mathbb{C}^{RL\times KL}$ and new vectors $\{\boldsymbol{z}_l^{n'}, \boldsymbol{z}_l^{n'} \in \mathbb{C}^{K\times 1}\}$. Then, we have
   \begin{align}
	\boldsymbol{Y}= \boldsymbol{H}'[\boldsymbol{z}^{1'}, \boldsymbol{z}^{2'}, \cdots, \boldsymbol{z}^{N'}] + \boldsymbol{V},  \label{eq:demmodel}
	\end{align} 
	 where
	\begin{align}
		\boldsymbol{z}^{n'} = [(\boldsymbol{z}_1^{n'})^T, (\boldsymbol{z}_2^{n'})^T, \cdots, (\boldsymbol{z}_L^{n'})^T]^T.
	\end{align}
	
	As discussed in Section \ref{sec:model}, data symbols $\{s_k^n\}$ are modulated with differential modulations, so $\boldsymbol{z}_1^{n'}$ can be expressed as $\boldsymbol{z}_1^{n'} = \boldsymbol{z}_1^{(n-1)'}\cdot \boldsymbol{s}^n$,
	where $\boldsymbol{s}^n\triangleq \{s_1^n, s_2^n, \cdots, s_K^n\}^T$ and $p(s_k^n) = \sum_{q\in \mathcal{A}}\frac{1}{|\mathcal{A}|}\delta(s_k^n-q)$. For notation simplification, \rev{we still use $\boldsymbol{H}$, $\boldsymbol{z}^{n}$ and $\boldsymbol{z}_l^n$ instead of  $\boldsymbol{H}'$, $\boldsymbol{z}^{n'}$ and $\boldsymbol{z}_l^{n'}$ in this subsection}. 
	
	Similar to the operations on \eqref{eq:Hmodel}, we firstly perform a unitary transformation on \eqref{eq:demmodel} and obtain the joint conditional probability of variables  $\{\boldsymbol{h}_l^n\}$, $\{\boldsymbol{z}_l^n\}$, $\{\boldsymbol{s}^n\}$ and $\gamma$, i.e.,
	\begin{align}
		&p(\{\boldsymbol{h}_l^n\}, \{\boldsymbol{z}_l^n\}, \{\boldsymbol{s}^n\},\gamma | \{\overline{\boldsymbol{y}}_l^n\}) \nonumber \\
		=& \prod\nolimits_{n,l}p(\overline{\boldsymbol{y}}_l^n|\boldsymbol{h}_l^n,\gamma)p(\boldsymbol{h}_l^n|\boldsymbol{z}_l^n)p(\boldsymbol{z}_l^n|\{\boldsymbol{s}^n\}, \boldsymbol{z}_l^{n-1})p(\boldsymbol{s}^n)p(\gamma) \nonumber \\
		\triangleq & \prod\nolimits_{n,l}f_{\overline{\boldsymbol{y}}_l^n}(\boldsymbol{h}_l^n,\gamma)f_{\boldsymbol{h}_l^n}(\boldsymbol{z}_l^n)f_{\delta_l^n}(\boldsymbol{z}_l^n,\{\boldsymbol{s}^n\}, \boldsymbol{z}_l^{n-1})f_{\boldsymbol{s}^n}(\boldsymbol{s}^n)f_{\gamma}(\gamma) \label{eq:comfactor}
	\end{align}
	The  factor graph representation of \eqref{eq:comfactor} is depicted in Fig. 3. The probability densities and the corresponding factors are listed in Table \uppercase\expandafter{\romannumeral2}. Next, we elaborate how to get the beliefs of the transmitted symbols using message passing in this graph, based on which \rev{the estimates of symbols can be obtained.} 
	
	\begin{table}[htb]
		\centering
		\renewcommand\arraystretch{1.2}
		\caption{Factors and distributions in (\ref{eq:comfactor}).}\label{tab:factor}
		\begin{tabular}{ccc}
			\hline
			Factor & Distribution & Function  \\
			\hline
			$f_{\overline{\boldsymbol{y}}_l^n}$&     $p(\overline{\boldsymbol{y}}_l^n|\boldsymbol{h}_l^n,\gamma)$ & $\mathcal{CN}(\overline{y}_l^n; \boldsymbol{h}_l^n, \gamma^{-1}\boldsymbol{I})$ \\	
			$f_{\boldsymbol{h}_l^n}$ & $p(\boldsymbol{h}_l^n|\boldsymbol{z}_l^n)$ & $\delta(\boldsymbol{h}_l^n - \boldsymbol{\Phi}_l\boldsymbol{z}_l^n)$\\
			$f_{\delta_l^n}$ & $p(\boldsymbol{z}_l^n|\{\boldsymbol{s}^n\}, \boldsymbol{z}_l^{n-1})$ & $\delta(\boldsymbol{z}^n - \boldsymbol{z}^{n-1}\cdot \boldsymbol{s}^{n-1})$\\
			$f_{\boldsymbol{s}^n}$ & $p(\boldsymbol{s}^n)$ & $\prod_{k}\sum_{q\in \mathcal{A}}\frac{1}{|\mathcal{A}|}\delta(s_k^n-q)$ \\
			$f_{\gamma}$ & $p(\gamma)$ & $\gamma^{-1}$ \\				
			\hline
		\end{tabular}
	\end{table}						

	\subsubsection{ Forward Message Passing}
	
	In Fig.\ref{fig:stage2}, the message from variable node $\boldsymbol{z}_l^n, 1\leq l \leq L, 2 \leq n < N$ to function node $f_{\delta_l^n}$ can be expressed as
	\begin{align}
		n_{\boldsymbol{z}_l^n \rightarrow f_{\delta_l^n}}(\boldsymbol{z}_l^n) &=  m_{f_{\boldsymbol{h}_l^n}\rightarrow \boldsymbol{z}_l^n}(\boldsymbol{z}_l^n)m_{f_{\delta_l^{n+1}}\rightarrow \boldsymbol{z}_l^n}(\boldsymbol{z}_l^n) \nonumber \\
		&\propto \mathcal{CN}\left(\boldsymbol{z}_l^n; \rarrow{\boldsymbol{z}}_l^n, \text{diag}(\rarrow{\boldsymbol{\nu}}_{\boldsymbol{z}_{l}^n})\right) 
	\end{align}
	with\rev{
	\begin{align}
		\rarrow{\boldsymbol{\nu}}_{\boldsymbol{z}_{l}^n} =& \left({1}./{\boldsymbol{\nu}_{\boldsymbol{q}_l^n}} + {1}./{\larrow{\boldsymbol{\nu}}_{\boldsymbol{z}_{l}^{n, n+1}}}\right)^{-1}, \label{eq:demod_rvz} \\
		\rarrow{\boldsymbol{z}}_{l}^n =& \rarrow{\boldsymbol{\nu}}_{\boldsymbol{z}_{l}^n} \cdot \left({\boldsymbol{q}_l^n}./{\boldsymbol{\nu}_{\boldsymbol{q}_l^n}} + {\larrow{\boldsymbol{z}}_{l}^{n, n+1}}./{\larrow{\boldsymbol{\nu}}_{\boldsymbol{z}_{l}^{n, n+1}}}\right) ,\label{eq:demod_rz}
	\end{align}
}
	where $\boldsymbol{\nu}_{\boldsymbol{q}_l^n}$ and $\boldsymbol{\nu}_{\boldsymbol{q}_n^l}$ are respectively given in \eqref{eq:vqnl} and \eqref{eq:qhatnl},
	 $\larrow{\boldsymbol{z}}_{l}^{n, n+1}$ and  $\larrow{\boldsymbol{\nu}}_{\boldsymbol{z}_{l}^{n, n+1}}$ denote the mean and variance of the message $m_{f_{\delta_l^{n+1}}\rightarrow \boldsymbol{z}_l^n}(\boldsymbol{z}_l^n)$, the computation of which is delayed to \eqref{eq:fdeltan2zn1}. Meanwhile, we can derive the message from variable node $\boldsymbol{z}_l^{n-1}$ to function node $f_{\delta_l^n}$, i.e., 
	\begin{align}
		&n_{\boldsymbol{z}_l^{n-1} \rightarrow f_{\delta_l^n}}(\boldsymbol{z}_l^{n-1}) \nonumber \\
		&= m_{f_{\boldsymbol{h}_l^{n-1}}\rightarrow \boldsymbol{z}_l^{n-1}}(\boldsymbol{z}_l^{n-1})m_{f_{\delta_l^{n-1}}\rightarrow \boldsymbol{z}_l^{n-1}}(\boldsymbol{z}_l^{n-1}) \nonumber \\
		&\propto \mathcal{CN}\left(\boldsymbol{z}_l^{n-1}; \rarrow{\boldsymbol{z}}_{l}^{n-1,n}, \text{diag}(\rarrow{\boldsymbol{\nu}}_{\boldsymbol{z}_{l}^{n-1,n}})\right), \label{eq:zln12deltaln}
	\end{align}
	where $m_{f_{\boldsymbol{h}_l^{n-1}}\rightarrow \boldsymbol{z}_l^{n-1}}(\boldsymbol{z}_l^{n-1})$ and
	$m_{f_{\delta_l^{n-1}}\rightarrow \boldsymbol{z}_l^{n-1}}(\boldsymbol{z}_l^{n-1})$ can be simply derived by  replacing the superscript $n$ of \eqref{eq:fh2z} and \eqref{eq:fdeltaln2zln} with $n-1$, respectively. In addition, $m_{f_{\delta_l^{n-1}}\rightarrow \boldsymbol{z}_l^{n-1}}(\boldsymbol{z}_l^{n-1})$ is approximated to be a complex Gaussian distribution with EP, and its computation is delayed to \eqref{eq:epapproximate}. Hence, $\rarrow{\boldsymbol{z}}_{l}^{n-1,n}$ and $\rarrow{\boldsymbol{\nu}}_{\boldsymbol{z}_{l}^{n-1,n}}$ can be further represented as
	\begin{align}
		\rarrow{\boldsymbol{\nu}}_{\boldsymbol{z}_{l}^{n-1,n}}=& \boldsymbol{1}./(\boldsymbol{1}./\boldsymbol{\nu}_{\boldsymbol{q}_l^{n-1}} + \boldsymbol{1} ./ \larrow{\boldsymbol{\nu}}_{\boldsymbol{z}_l^{n-1}}), \label{eq:demod_rvzln1n} \\
		\rarrow{\boldsymbol{z}}_{l}^{n-1,n} =& \rarrow{\boldsymbol{\nu}}_{\boldsymbol{z}_{l}^{n-1,n}}\cdot(\boldsymbol{q}_l^{n-1}./\boldsymbol{\nu}_{\boldsymbol{q}_l^{n-1}} + \larrow{\boldsymbol{z}}_{l}^{n-1} ./ \larrow{\boldsymbol{\nu}}_{\boldsymbol{z}_l^{n-1}}).\label{eq:demod_rzln1n}
	\end{align}
	
	Next, the message from function node $f_{\delta_l^n}$ to variable node $\boldsymbol{s}^n$ can be computed with BP, i.e.,
	\begin{align}
		&m_{f_{\delta_l^n} \rightarrow \boldsymbol{s}^n}(\boldsymbol{s}^n) \nonumber\\ &= \int f_{\delta_l^n}(\boldsymbol{z}_l^n, \boldsymbol{z}_l^{n-1}, \boldsymbol{s}^n)	n_{\boldsymbol{z}_l^n \rightarrow f_{\delta_l^n}}(\boldsymbol{z}_l^n) \nonumber\\
		&  \hspace{3cm}\times n_{\boldsymbol{z}_l^{n-1} \rightarrow f_{\delta_l^n}}(\boldsymbol{z}_l^{n-1})d\boldsymbol{z}_l^nd\boldsymbol{z}_l^{n-1} \nonumber \\
		&\propto \mathcal{CN}\left(\rarrow{\boldsymbol{z}}_{l}^n./\boldsymbol{s}^n; \rarrow{\boldsymbol{z}}_{l}^{n-1,n}, diag\left(\rarrow{\boldsymbol{\nu}}_{\boldsymbol{z}_{l}^n}./|\boldsymbol{s}^n|^2 +\rarrow{\boldsymbol{\nu}}_{\boldsymbol{z}_{l}^{n-1,n}}\right )\right), 
	\end{align}
	where $f_{\delta_l^n}(\boldsymbol{z}_l^n, \boldsymbol{z}_l^{n-1}, \boldsymbol{s}^n)\triangleq \delta(\boldsymbol{z}_l^n - \boldsymbol{z}_l^{n-1}\cdot \boldsymbol{s}^n)$. \rev{Meanwhile, we can further obtain that 
	\begin{align}
		&m_{f_{\delta_l^n} \rightarrow s_k^n}(s_k^n) \nonumber \\
		&\propto \mathcal{CN} (\rarrow{z}_{k,l}^n/s_k^n;\rarrow{z}_{k,l}^{n-1,n},\rarrow{\nu}_{z_{k,l}^n}/|s_k^n|^2 + \rarrow{\nu}_{z_{k,l}^{n-1,n}}),
	\end{align}	
	where the subscript $k$ denotes that the scalar $a_k$ is the $k$-th element of the vector $\boldsymbol{a}$.
   }

	\rev{With $\{m_{f_{\delta_l^n} \rightarrow s_k^n}(s_k^n), 1\leq l \leq L\}$ and the prior $f_{s_k^n}(s_k^n) \triangleq \sum_{q\in \mathcal{A}}\frac{1}{|\mathcal{A}|}\delta(s_k^n - q)$,  the belief of $s_k^n$ can be expressed as
	\begin{align}
		b(s_k^n) =& f_{s_k^n}(s_k^n)\prod\nolimits_{l=1}^L m_{f_{\delta_l^n} \rightarrow s_k^n}(s_k^n) \nonumber \\
		=& \sum\nolimits_{q\in \mathcal{A}} \psi_{k,q}^n\delta(s_k^n - q) \label{eq:demod_bsn}
	\end{align}
	where $\psi_{k,q}^n$ is given as
	\begin{align}
		\psi_{k,q}^n = \frac{\prod\limits_{l=1}^L \mathcal{CN}\left(z_{k,l}^n; q\rarrow{z}_{k,l}^{n-1,n}, \rarrow{\nu}_{z_{k,l}^n} +|q|^2\rarrow{\nu}_{z_{k,l}^{n-1,n}}\right)}{
		\sum\limits_{q'\in \mathcal{A}}\prod\limits_{l=1}^L \mathcal{CN}\left(z_{k,l}^n; q'\rarrow{z}_{k,l}^{n-1,n}, \rarrow{\nu}_{z_{k,l}^n} +|q'|^2\rarrow{\nu}_{z_{k,l}^{n-1,n}}\right)
		} \label{eq:psi}
	\end{align}
}
		\subsubsection {Backward Message Passing}
	
	The message $n_{\boldsymbol{s}^n\rightarrow f_{\delta_l^n}}(\boldsymbol{s}^n)$  from variable node $\boldsymbol{s}^n$ to function node $f_{\delta_l^n}$ can be computed as
	\begin{align}
	 n_{\boldsymbol{s}^n\rightarrow f_{\delta_l^n}}(\boldsymbol{s}^n) =& f_{\boldsymbol{s}^n}(\boldsymbol{s}^n)\prod_{l'\neq l}m_{f_{\delta_{l'}^n} \rightarrow \boldsymbol{s}^n}(\boldsymbol{s}^n) \nonumber\\
	 \triangleq & \prod\nolimits_{k=1}^K\sum\nolimits_{q\in \mathcal{A}} \xi_{k,l,q}^n\delta(s_k^n - q),
	\end{align}
	where 
		\begin{align}
		\xi_{k,l,q}^n = \frac{\prod\limits_{l'\neq l} \mathcal{CN}\left(z_{k,l'}^n; q\rarrow{z}_{k,l'}^{n-1,n}, \rarrow{\nu}_{z_{k,l'}^n} +|q|^2\rarrow{\nu}_{z_{k,l'}^{n-1,n}}\right)}{
			\sum\limits_{q'\in \mathcal{A}}\prod\limits_{l'\neq l} \mathcal{CN}\left(z_{k,l'}^n; q'\rarrow{z}_{k,l'}^{n-1,n}, \rarrow{\nu}_{z_{k,l'}^n} +|q'|^2\rarrow{\nu}_{z_{k,l'}^{n-1,n}}\right)
		}.
	\end{align}
	
	With the message $ n_{\boldsymbol{s}^n\rightarrow f_{\delta_l^n}}(\boldsymbol{s}^n)$ and $n_{z_l^{n-1} \rightarrow f_{\delta_l^n}}(\boldsymbol{z}_l^{n-1})$, the message from function node $f_{\delta_l^n}$ to variable node $\boldsymbol{z}_l^n$ can be expressed as
	\begin{align}
	&	m_{f_{\delta_l^n\rightarrow \boldsymbol{z}_l^n}}(\boldsymbol{z}_l^n) \nonumber\\ &= \int f_{\delta_l^n}(\boldsymbol{z}_l^n, \boldsymbol{z}_l^{n-1}, \boldsymbol{s}^n)n_{\boldsymbol{s}^n\rightarrow f_{\delta_l^n}}(\boldsymbol{s}^n) \nonumber\\ &\hspace{3cm} \times n_{\boldsymbol{z}_l^{n-1} \rightarrow f_{\delta_l^n}}(\boldsymbol{z}_l^{n-1})d\boldsymbol{s}^nd\boldsymbol{z}^{n-1} \nonumber \\
	&= \prod\nolimits_k \sum\nolimits_{q\in \mathcal{A}}\xi_{k,q}^n|q|^2\mathcal{CN}(z_{l,k}^n; q\rarrow{z}_{k,l}^{n-1,n}, |q|^2\rarrow{\nu}_{z_{k,l}^{n-1,n}}). \label{eq:fdeltaln2zln}
	\end{align}
	Note that $m_{f_{\delta_l^n\rightarrow \boldsymbol{z}_l^n}}(\boldsymbol{z}_l^n)$ is discrete, making the following message computations intractable. To sovle this problem, we can approximate it to be Gaussian, i.e.,
	\begin{align}
		m_{f_{\delta_l^n\rightarrow \boldsymbol{z}_l^n}}(\boldsymbol{z}_l^n) \approx \mathcal{CN}(\boldsymbol{z}_l^n; \larrow{\boldsymbol{z}}_{l}^n, diag(\larrow{\boldsymbol{\nu}}_{\boldsymbol{z}_{l}^n})), \label{eq:epapproximate}
	\end{align}
	where the $k$-th elements of $\larrow{\boldsymbol{z}}_{l}^n$ and $\larrow{\boldsymbol{\nu}}_{\boldsymbol{z}_{l}^n}$ (i.e., $\larrow{z}_{k, l}^n$ and $\larrow{\nu}_{z_{k,l}^n}$) can be updated with
	\begin{align}
		 \larrow{z}_{k,l}^n =& \sum\nolimits_{q\in \mathcal{A}}\alpha_{k,l,q}^nq\rarrow{z}_{k,l}^{n-1,n}, \label{eq:demod_lzkln} \\
		 \larrow{\nu}_{z_{k,l}^n} =&  \sum\nolimits_{q\in \mathcal{A}}\alpha_{k,l,q}^n\left(|q\rarrow{z}_{k,l}^{n-1,n}|^2 + |q|^2\rarrow{\nu}_{z_{k,l}^{n-1,n}}\right), \\
		 \alpha_{k,l,q}^n =&  \frac{\xi_{k,l,q}^n|q|^2}{\sum_{q'\in\mathcal{A}}\xi_{k,q'}^n|q'|^2}. \label{eq:demod_lvzkln}
	\end{align}
	
	Similar to the message $m_{f_{\delta_l^n\rightarrow \boldsymbol{z}_l^n}}(\boldsymbol{z}_l^n)$, the message from function node $f_{\delta_l^n}$
	to variable node $\boldsymbol{z}_l^{n-1}$ can be expressed as
	\begin{align}
	&	m_{f_{\delta_l^n\rightarrow \boldsymbol{z}_l^{n-1}}}(\boldsymbol{z}_l^{n-1}) \nonumber\\ &= \int f_{\delta_l^n}(\boldsymbol{z}_l^n, \boldsymbol{z}_l^{n-1}, \boldsymbol{s}^n)n_{\boldsymbol{s}^n\rightarrow f_{\delta_l^n}}(\boldsymbol{s}^n)  n_{\boldsymbol{z}_l^{n} \rightarrow f_{\delta_l^n}}(\boldsymbol{z}_l^{n})d\boldsymbol{s}^nd\boldsymbol{z}^{n} \nonumber \\
	&\approx \mathcal{CN}(\boldsymbol{z}_l^{n-1}; \larrow{\boldsymbol{z}}_{l}^{n-1,n}, \text{diag}(\larrow{\boldsymbol{\nu}}_{\boldsymbol{z}_{l}^{n-1,n}})) \label{eq:fdeltan2zn1}
\end{align}
	with the $k$-th elements of $\larrow{\boldsymbol{z}}_{l}^{n-1}$ and $\larrow{\boldsymbol{\nu}}_{\boldsymbol{z}_{l}^{n-1}}$ (i.e., $\larrow{z}_{k, l}^{n-1}$ and $\larrow{\nu}_{z_{k,l}^{n-1}}$) given as
	\begin{align}
	\larrow{z}_{k,l}^{n-1,n} =& \sum\nolimits_{q\in \mathcal{A}}\rev{\alpha_{k,l,q}^n}q\rarrow{z}_{k,l}^{n}, 
	\label{eq:demod_lzkln1n} \\
	\larrow{\nu}_{z_{k,l}^{n-1,n}} =&  \sum\nolimits_{q\in \mathcal{A}}\rev{\alpha_{k,l,q}^n}\left(|q\rarrow{z}_{k,l}^{n}|^2 + |q|^2\rarrow{\nu}_{z_{k,l}^{n}}\right), \label{eq:demod_lvzkln1n}
\end{align}
	Similarly, we can obtain that
	\begin{align}
		m_{f_{\delta_l^{n+1}\rightarrow \boldsymbol{z}_l^{n}}}(\boldsymbol{z}_l^{n}) \approx 
		\mathcal{CN}(\boldsymbol{z}_l^{n}; \larrow{\boldsymbol{z}}_{l}^{n, n+1}, diag(\larrow{\boldsymbol{\nu}}_{\boldsymbol{z}_{l}^{n,n+1}})).
	\end{align}
	Next, along with the messages $m_{f_{\delta_l^n\rightarrow \boldsymbol{z}_l^n}}(\boldsymbol{z}_l^n)$ and $m_{f_{\boldsymbol{h}_l^n}\rightarrow \boldsymbol{z}_l^n}(\boldsymbol{z}_l^n)$, which are respectively given in \eqref{eq:epapproximate} and \eqref{eq:fh2z}, the belief of $\boldsymbol{z}_l^n$ can be updated as
	\begin{align}
		b(\boldsymbol{z}_l^n) =& m_{f_{\delta_l^{n+1}\rightarrow \boldsymbol{z}_l^{n}}}(\boldsymbol{z}_l^{n})m_{f_{\delta_l^n\rightarrow \boldsymbol{z}_l^n}}(\boldsymbol{z}_l^n)m_{f_{\boldsymbol{h}_l^n}\rightarrow \boldsymbol{z}_l^n}(\boldsymbol{z}_l^n) \nonumber \\
		\propto& \mathcal{CN}(\boldsymbol{z}_l^n; \hat{\boldsymbol{z}}_l^n, \boldsymbol{\nu}_{\boldsymbol{z}_l^n}),
	\end{align}
	where
	\begin{align}
		\boldsymbol{\nu}_{\boldsymbol{z}_l^n} =& \boldsymbol{1}./\left(\boldsymbol{1}./\boldsymbol{\nu}_{\boldsymbol{q}_n^l} + \boldsymbol{1}./\larrow{\boldsymbol{\nu}}_{\boldsymbol{z}_{l}^{n,n+1}} + \boldsymbol{1}./\larrow{\boldsymbol{\nu}}_{\boldsymbol{z}_{l}^n} \right) \label{eq:demod_mz} \\
		\hat{\boldsymbol{z}}_l^n =& \boldsymbol{\nu}_{\boldsymbol{z}_l^n} \left(\boldsymbol{q}_l^n./\boldsymbol{\nu}_{\boldsymbol{q}_n^l} + \larrow{\boldsymbol{z}}_{l}^{n,n+1}./\larrow{\boldsymbol{\nu}}_{\boldsymbol{z}_{l}^{n,n+1}} + \larrow{\boldsymbol{z}}_{l}^n./\larrow{\boldsymbol{\nu}}_{\boldsymbol{z}_{l}^n} \right) \label{eq:demod_vz}
	\end{align}

	It is noted that the subsequent backward message computations, such as $m_{f_{\boldsymbol{h}_l^n\rightarrow \boldsymbol{h}_l^n}}(\boldsymbol{h}_l^n)$, are the same with those in Section \ref{subsec:mplocalization}, thus we omit them. The backward and forward message passing is performed alternatively, and we also terminate the iteration when normalized squared difference of the {estimated vectors $\{\boldsymbol{z}_l^n\}$} between two consecutive iterations is less than a threshold $\xi$. \rev{Finally, the decision on the transmitted symbol $s_k^n$ can be made, i.e.,
		\begin{align}
			\hat{s}_k^n = \mathop{\arg\max}\limits_{q} \psi_{k,q}^n, ~q \in \mathcal{A}, \label{eq:harddec}
	\end{align}
	where $\psi_{k,q}^n$ is given in \eqref{eq:psi}.
   } 

	\begin{algorithm}
		\setstretch{1.25}
		\caption{UAMP-Based JCL Algorithm}
		\textbf{Initialize}: Unitary transform on \eqref{eq:Hmodel}. $\hat{\lambda}_m = 1$, $\boldsymbol{q}_l^n=\boldsymbol{0}$, $\boldsymbol{\nu}_{\boldsymbol{q}_l^n}=\boldsymbol{1}$, $\boldsymbol{e}_l^n=\boldsymbol{0}$ and $\epsilon'=1$, $\hat{\gamma}=1$, $\xi=10^{-5}$.\\
		\textbf{Repeat}
		\begin{algorithmic}[1]		
			\STATE $\forall l, n$ : update $\boldsymbol{\nu}_{\boldsymbol{z}_l^n}$ and $\hat{\boldsymbol{z}}_l^n$ with \eqref{eq:vznl} and \eqref{eq:zhatnl}
			\STATE $\forall l, n$ : update $\boldsymbol{\nu}_{\boldsymbol{p}_l^n}$ and $\boldsymbol{p}_l^n$ with \eqref{eq:vpnl} and \eqref{eq:phatnl}	
			\STATE $\forall l, n$ : update $\boldsymbol{\nu}_{\boldsymbol{h}_l^n}$ and $\hat{\boldsymbol{h}_l^n}$ with \eqref{eq:vhnl} and \eqref{eq:hhatnl}
			\STATE update noise precision $\hat{\gamma}$ with \eqref{eq:gammaest}
			\STATE $\forall l, n$ : update $\boldsymbol{\nu}_{\boldsymbol{e}_l^n}$ and $\hat{\boldsymbol{e}_l^n}$ with \eqref{eq:veln} and \eqref{eq:eln}
			\STATE $\forall l, n$ : update $\boldsymbol{\nu}_{\boldsymbol{q}_l^n}$ and $\hat{\boldsymbol{q}_l^n}$ with \eqref{eq:vqnl} and \eqref{eq:qhatnl}
			\STATE $\forall l, n$ : update $\boldsymbol{\nu}_{\boldsymbol{z}_l^n}$ and $\hat{\boldsymbol{z}}_l^n$ with \eqref{eq:vznl} and \eqref{eq:zhatnl}
			\STATE $\forall m$ : update $\hat{\lambda}_m$ with \eqref{eq:lambdam}	
			\STATE  update $\epsilon'$ with \eqref{eq:epsilon}
		\end{algorithmic}
		\textbf{Until terminated}
		\begin{enumerate} 
			\addtocounter{enumi}{+9}
			\renewcommand{\labelenumi}{\footnotesize \theenumi:}
			\setlength{\itemindent}{-0.5em}
			\item Perform unitary transformation on \eqref{eq:demmodel} 
		\end{enumerate}	 
		\textbf{Repeat}
		 \begin{enumerate} 
			\addtocounter{enumi}{+10}
			\renewcommand{\labelenumi}{\footnotesize \theenumi:}
			\setlength{\itemindent}{-0.5em}
			\item $\forall l, n$: update $\rarrow{\boldsymbol{\nu}}_{\boldsymbol{z}_l^n}$ and $\rarrow{\boldsymbol{z}}_l^n$ with \eqref{eq:demod_rvz} and \eqref{eq:demod_rz}
			\item $\forall l, n>1$: update $\rarrow{\boldsymbol{\nu}}_{\boldsymbol{z}_l^{n-1,n}}$ and $\rarrow{\boldsymbol{z}}_l^{n-1,n}$ with \eqref{eq:demod_rvzln1n} and \eqref{eq:demod_rzln1n}
			\item $\forall n>1$: update the belief of $\boldsymbol{s}^n$ with \eqref{eq:demod_bsn}
			\item $\forall k,l,n$: update $\larrow{\boldsymbol{z}}_{k,l}^n$ and $\larrow{\nu}_{z_{k,l}^n}$ with \eqref{eq:demod_lzkln} and \eqref{eq:demod_lvzkln}
			\item $\forall k,l,n>1$: update $\larrow{z}_{k,l}^{n-1,n}$ and $\larrow{\nu}_{z_{k,l}^{n-1,n}}$ with \eqref{eq:demod_lzkln1n} and \eqref{eq:demod_lvzkln1n}
			\item  $\forall l,n$: update $\boldsymbol{\nu}_{\boldsymbol{z}_l^n}$ and $\hat{\boldsymbol{z}}_l^n$ with \eqref{eq:demod_mz} and \eqref{eq:demod_vz}
			\item $\forall l, n$: update $\boldsymbol{\nu}_{\boldsymbol{p}_l^n}$ and $\boldsymbol{p}_l^n$ with \eqref{eq:vpnl} and \eqref{eq:phatnl}	
			\item $\forall l, n$: update $\boldsymbol{\nu}_{\boldsymbol{h}_l^n}$ and $\hat{\boldsymbol{h}_l^n}$ with \eqref{eq:vhnl} and \eqref{eq:hhatnl}
			\item update noise precision $\hat{\gamma}$ with \eqref{eq:gammaest}
			\item $\forall l, n$: update $\boldsymbol{\nu}_{\boldsymbol{e}_l^n}$ and $\hat{\boldsymbol{e}_l^n}$ with \eqref{eq:veln} and \eqref{eq:eln}
			\item $\forall l, n$: update $\boldsymbol{\nu}_{\boldsymbol{q}_l^n}$ and $\hat{\boldsymbol{q}_l^n}$ with \eqref{eq:vqnl} and \eqref{eq:qhatnl}
		\end{enumerate}	 
	   \textbf{Until terminated}
	   \begin{enumerate} 
	   	\addtocounter{enumi}{21}
	   	\renewcommand{\labelenumi}{\footnotesize \theenumi:}
	   	\setlength{\itemindent}{-0.5em}
	   	\item $\forall k,n $: \rev{determine the estimate of $s_k^n$ with \eqref{eq:harddec}}
	   \end{enumerate}	 
		\label{algorithm:blockUTAMP}
	\end{algorithm}
	The proposed algorithm is summarized in Algorithm.\ref{algorithm:blockUTAMP}. It can be seen that the complexity of each iteration is dominated by the matrix-vector products, which is $\mathcal{O}(LRMN)$.	
	
	\section{Simulation Results}
	As illustrated in Fig.\ref{fig:sysmodel}, we consider a $20m \times 20m$  area served by 4 base stations, and the coordinate of its top-left corner is set to (0,0). Users are randomly located in the area. The BSs are assumed 5 meters away from the boundary of area, and are placed at (-5m, 10m), (10m, 25m), (25m, 10m), and (10m, -5m).  The grid interval in both directions is 1m. The modulation scheme employed is $\pi/4$ DQPSK. The threshold $\xi$ is set to $10^{-5}$ to terminate the iterative process of the algorithm. The number of antenna $R$ at the BSs is set to 128. The signal to noise ratio (SNR) is defined as $\text{SNR}=||(\overline{\boldsymbol{A}}~\overline{\boldsymbol{X}})||_2^2/(K\gamma^{-1})$.

	\rev{We evaluate the localization performance and bit error rate (BER) performance of the proposed method. The localization accuracy is defined as $C^K/C$, where C denotes the total number of trials and $C^K$ denotes the number of trails where the locations of all $K$ users} are identified correctly.   
	 \begin{figure}[htbp]
		\centering
		\includegraphics[width=\columnwidth]{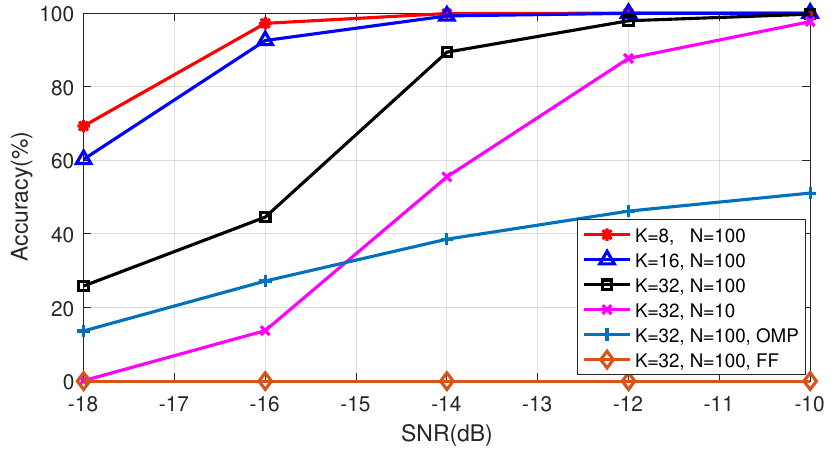} 			
		\caption{{Localization accuracy comparison of the algorithms.}} 
		\label{fig:localizationres}
	\end{figure} 
	
	In Fig.\ref{fig:localizationres}, we set $N=100$ and the localization accuracy of the proposed algorithm is shown with different number of users $K$, where the case with $N=10$ and $K=32$ is also investigated. We can see that the proposed method exhibits high localization accuracy, especially when SNR is larger than -10dB, the localization accuracy is almost $100\%$. In addition, we also examine the impact of modelling error by assuming that the users are all in FF, and the performance is indicated by the curve `K=32, N=100, FF', where we can see that the system simply does not work. The orthogonal matching pursuit (OMP) algorithm is also used for comparison, and results in Fig. 4 shows that our proposed algorithm delivers superior performance. 
	
	 In Fig.\ref{fig:berres}, $N$ is still set to 100 and the BER of the system with different number of users is shown. We can see that the proposed algorithm delivers good BER performance, and as expected, with the decrease of the user number, the BER performance of the system slightly improves. In addition, we also investigate the BER performance of OMP with $K=8$, where once the estimate of $\boldsymbol{z}_l^n$ is obtained, which is denoted by $\hat{\boldsymbol{z}}_l^n$,  \rev{$\hat{s}_k^n$ is determined with $\hat{s}_k^n=\frac{1}{L}\sum_{l=1}^L(\hat{z}_{k,l}^{n}/\hat{z}_{k,l}^{n-1})$ followed by the hard decision on $\mathcal{A}$}. The comparison shows that the proposed algorithm significantly outperforms OMP. 
	 
	 As the noise variance estimation is integrated in the proposed algorithm, we examine its estimation accuracy with different SNRs in Fig.\ref{fig:noiseres}. We can see that the proposed algorithm can achieve accurate noise variance estimation, especially when the SNR is relatively high.

	\begin{figure}[htbp]
		\centering
		\includegraphics[width=\columnwidth]{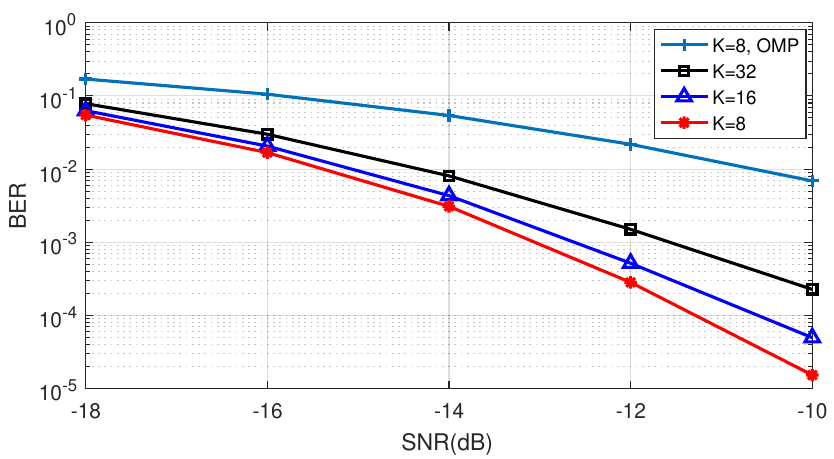} 			
		\caption{{BER comparisons versus SNR}} 
		\label{fig:berres}
	\end{figure}

	\begin{figure}[htbp]
		\centering
		\includegraphics[width=\columnwidth]{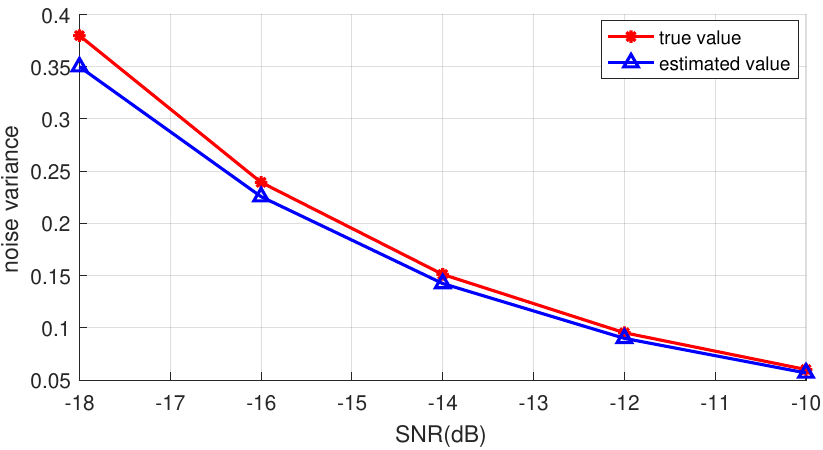} 			
		\caption{{Comparison between the estimated noise variance and true noise variance versus SNR}} 
		\label{fig:noiseres}
	\end{figure}

\section {Conclusions}
In this paper, we have investigated the problem of near field joint uplink communication and localization in a mobile network. We formulate the joint localization and communication problem as a sparse signal recovery problem, and leveraging UAMP, effective message passing algorithms are developed for joint localization and signal detection. Simulation results demonstrate the superior performance of the proposed algorithm.

	\vspace{-0.1cm}

	\bibliographystyle{IEEEtran}
	\bibliography{IEEEabrv,bibliography}

\begin{thebibliography}{10}
\providecommand{\url}[1]{#1}
\csname url@samestyle\endcsname
\providecommand{\newblock}{\relax}
\providecommand{\bibinfo}[2]{#2}
\providecommand{\BIBentrySTDinterwordspacing}{\spaceskip=0pt\relax}
\providecommand{\BIBentryALTinterwordstretchfactor}{4}
\providecommand{\BIBentryALTinterwordspacing}{\spaceskip=\fontdimen2\font plus
\BIBentryALTinterwordstretchfactor\fontdimen3\font minus
  \fontdimen4\font\relax}
\providecommand{\BIBforeignlanguage}[2]{{%
\expandafter\ifx\csname l@#1\endcsname\relax
\typeout{** WARNING: IEEEtran.bst: No hyphenation pattern has been}%
\typeout{** loaded for the language `#1'. Using the pattern for}%
\typeout{** the default language instead.}%
\else
\language=\csname l@#1\endcsname
\fi
#2}}
\providecommand{\BIBdecl}{\relax}
\BIBdecl

\bibitem{cui2023}
M.~Cui, Z.~Wu, Y.~Lu, X.~Wei, and L.~Dai, ``Near-field mimo communications for
  6g: Fundamentals, challenges, potentials, and future directions,'' \emph{IEEE
  Communications Magazine}, vol.~61, no.~1, pp. 40--46, 2023.

\bibitem{9508850}
A.~Guerra, F.~Guidi, and Dardari., ``{Near-Field Tracking With Large Antenna
  Arrays: Fundamental Limits and Practical Algorithms},'' \emph{IEEE
  Transactions on Signal Processing}, vol.~69, pp. 5723--5738, 2021.

\bibitem{liu2023}
Y.~Lu and L.~Dai, ``Near-field channel estimation in mixed los/nlos
  environments for extremely large-scale mimo systems,'' \emph{IEEE
  Transactions on Communications}, vol.~71, no.~6, pp. 3694--3707, 2023.

\bibitem{nftechrxiv}
S.~Hu, M.~C. Ilter, and H.~Wang, ``Near-field beamforming for large intelligent
  surfaces,'' in \emph{2022 IEEE 33rd Annual International Symposium on
  Personal, Indoor and Mobile Radio Communications (PIMRC)}, 2022, pp.
  1367--1373.

\bibitem{wei2022}
X.~Wei and L.~Dai, ``Channel estimation for extremely large-scale massive mimo:
  Far-field, near-field, or hybrid-field?'' \emph{IEEE Communications Letters},
  vol.~26, no.~1, pp. 177--181, 2022.

\bibitem{wang2023}
Z.~Wang, X.~Mu, and Y.~Liu, ``Near-field integrated sensing and
  communications,'' \emph{IEEE Communications Letters}, vol.~27, no.~8, pp.
  2048--2052, 2023.

\bibitem{andrewZhangConcurrent}
X.~Chen, Z.~Feng, Z.~Wei, J.~A. Zhang, X.~Yuan, and P.~Zhang, ``{Concurrent
  Downlink and Uplink Joint Communication and Sensing for 6G Networks},''
  \emph{IEEE Transactions on Vehicular Technology}, vol.~72, no.~6, pp.
  8175--8180, 2023.

\bibitem{andrewZhangMultiple}
X.~Chen, Z.~Feng, Z.~Wei, X.~Yuan, P.~Zhang, J.~Andrew~Zhang, and H.~Yang,
  ``{Multiple Signal Classification Based Joint Communication and Sensing
  System},'' \emph{IEEE Transactions on Wireless Communications}, pp. 1--1,
  2023.

\bibitem{32276}
R.~Roy and T.~Kailath, ``{ESPRIT-estimation of signal parameters via rotational
  invariance techniques},'' \emph{IEEE Transactions on Acoustics, Speech, and
  Signal Processing}, vol.~37, no.~7, pp. 984--995, 1989.

\bibitem{8736783}
B.~Friedlander, ``{Localization of Signals in the Near-Field of an Antenna
  Array},'' \emph{IEEE Transactions on Signal Processing}, vol.~67, no.~15, pp.
  3885--3893, 2019.

\bibitem{primerNFBeforming}
E.~Bjornson, O.~T. Demir, and L.~Sanguinetti, ``{A Primer on Near-Field
  Beamforming for Arrays and Reconfigurable Intelligent Surfaces},'' in
  \emph{2021 55th Asilomar Conference on Signals, Systems, and Computers},
  2021, pp. 105--112.

\bibitem{nfanalysis}
H.~Lu and Y.~Zeng, ``{Communicating With Extremely Large-Scale Array/Surface:
  Unified Modeling and Performance Analysis},'' \emph{IEEE Transactions on
  Wireless Communications}, vol.~21, no.~6, pp. 4039--4053, 2022.

\bibitem{9709801}
O.~Rinchi, A.~Elzanaty, and M.-S. Alouini, ``{Compressive Near-Field
  Localization for Multipath RIS-Aided Environments},'' \emph{IEEE
  Communications Letters}, vol.~26, no.~6, pp. 1268--1272, 2022.

\bibitem{Guo2015UtAMP}
\BIBentryALTinterwordspacing
Q.~Guo and J.~Xi, ``{Approximate Message Passing with Unitary
  Transformation},'' \emph{CoRR}, vol. abs/1504.04799, 2015. [Online].
  Available: \url{http://arxiv.org/abs/1504.04799}
\BIBentrySTDinterwordspacing

\bibitem{NFSource}
Z.~Zheng, M.~Fu, W.-Q. Wang, S.~Zhang, and Y.~Liao, ``Localization of mixed
  near-field and far-field sources using symmetric double-nested arrays,''
  \emph{IEEE Transactions on Antennas and Propagation}, vol.~67, no.~11, pp.
  7059--7070, 2019.

\bibitem{NFChEstPosition}
\BIBentryALTinterwordspacing
S.~J. J.~W. Yijin~Pan, Cunhua~Pan, ``{Joint Channel Estimation and Localization
  in the Near Field of RIS Enabled mmWave/subTHz Communications},''
  \emph{CoRR}, vol. abs/2208.11343, 2016. [Online]. Available:
  \url{https://arxiv.org/abs/2208.11343}
\BIBentrySTDinterwordspacing

\bibitem{Tipping2001Sparse}
M.~E. Tipping, ``{Sparse Bayesian Learning and the Relevance Vector Machine},''
  \emph{Journal of Machine Learning Research}, vol.~1, no.~3, pp. 211--244,
  2001.

\bibitem{ZHANG2017344}
\BIBentryALTinterwordspacing
C.~Zhang, Z.~Yuan, Z.~Wang, and Q.~Guo, ``{Low complexity sparse Bayesian
  learning using combined belief propagation and mean field with a stretched
  factor graph},'' \emph{Signal Processing}, vol. 131, pp. 344--349, 2017.
  [Online]. Available:
  \url{https://www.sciencedirect.com/science/article/pii/S0165168416302134}
\BIBentrySTDinterwordspacing

\bibitem{luomansbl}
M.~Luo, Q.~Guo, M.~Jin, Y.~C. Eldar, D.~Huang, and X.~Meng, ``{Unitary
  Approximate Message Passing for Sparse Bayesian Learning},'' \emph{IEEE
  Transactions on Signal Processing}, vol.~69, pp. 6023--6039, 2021.

\end{thebibliography}
\end{document}